\documentclass[prb,aps,amsmath,twocolumn,amssymb,floatfixng,showpacs,
superscriptaddress]{revtex4}

\usepackage[dvips]{graphics}
\usepackage{color}
\definecolor{dred}{rgb}{0,0.65,0.2}

\begin{document}

\title{\textcolor{dred}{Electric field induced localization phenomena 
in a ladder network with superlattice configuration: Effect of backbone
environment}}

\author{Paramita Dutta}

\email{paramita.dutta@saha.ac.in}

\affiliation{Condensed Matter Physics Division, Saha Institute of Nuclear 
Physics, Sector-I, Block-AF, Bidhannagar, Kolkata-700 064, India}

\author{Santanu K. Maiti}

\email{santanu.maiti@isical.ac.in}

\affiliation{Physics and Applied Mathematics Unit, Indian Statistical
Institute, 203 Barrackpore Trunk Road, Kolkata-700 108, India}

\author{S. N. Karmakar}

\email{sachindranath.karmakar@saha.ac.in}

\affiliation{Condensed Matter Physics Division, Saha Institute of Nuclear 
Physics, Sector-I, Block-AF, Bidhannagar, Kolkata-700 064, India}

\begin{abstract}

Electric field induced localization properties of a tight-binding ladder
network in presence of backbone sites are investigated. Based on Green's
function formalism we numerically calculate two-terminal transport 
together with density of states for different arrangements of atomic 
sites in the ladder and its backbone. Our results lead to a possibility
of getting multiple mobility edges which essentially plays a switching
action between a completely opaque to fully or partly conducting region
upon the variation of system Fermi energy, and thus, support in 
fabricating mesoscopic or DNA-based switching devices.

\end{abstract}

\pacs{73.23.-b, 71.30.+h, 71.23.An, 73.63.Rt}

\maketitle

\section{Introduction}

Manifestation of localization of single particle states in low-dimensional 
quantum systems has always been in limelight of research in condensed 
matter physics since its prediction. Studies on this topic was basically 
stimulated after the work of Anderson in 1958~\cite{anderson} which has 
become a milestone in material science. It is well set that exponentially 
localized states are the only allowed solutions for one-dimensional ($1$D) 
systems with random on-site potentials~\cite{anderson,sharma}. Apart from
the above mentioned fact, another kind of localization phenomenon, the 
so-called Wannier-Stark localization~\cite{wannier}, has also drawn
much attention of many physicists~\cite{borysowicz,zekri}. This 
type of localizing behavior is observed in $1$D systems subjected to an
external electric field, even in absence of any kind of disorder. Although 
the existence of localized surface states in crystals has been first 
suggested by Tamm~\cite{tamm}, on the basis of a special $1$D model proposed 
by James~\cite{james}, popularity of this phenomenon among scientists has 
been started to grow following the work of Wannier~\cite{wannier} which 
has won its way after the widespread application to the optical properties 
of quantum wells~\cite{harwit,levine1}. 

For both these two typical cases i.e., systems with random site energies 
and samples with external electric field, one cannot find the existence of
{\em mobility edge} separating the conducting states from the non-conducting
region, since all the energy levels are localized. Therefore, for these
models no long-range transport will be obtained. However, a large number
of physically relevant models are available which exhibit mobility edges
at some typical energy values. For example, a $1$D tight-binding (TB) chain
composed of two uncorrelated random atomic sites exhibits extended energy 
eigenstates for some specific electron energies over the entire material
under a suitable alignment of these random sites as originally explored 
by Dunlap {\em et al.}~\cite{dunlap}. In this case, a short range positional 
correlation among the atomic sites is established. On the other hand,
in quasi-periodic Aubry-Andre model~\cite{aubry}, where long range positional 
correlation is found to exist between the atomic sites in $1$D geometry, 
conducting energy levels are also noticed. All these special classes of 
materials, the so-called correlated disordered materials, have provided 
a new turn in the localization phenomena.

Although few attempts~\cite{eco,das,rolf,sch,san1,san2,san3} have been made 
to explore the existence of mobility edges in some one- and two-dimensional 
($2$D) systems, a detailed analysis of it is still missing which essentially 
motivates us to investigate this phenomenon with a renewed interest. In this 
work we investigate localization properties of a tight-binding ladder 
network, constructed by coupling two superlattice chains laterally
(see Fig.1), in presence of an external electric field. Incidentally, many 
interesting and novel features of electronic properties have already been 
examined in different types of $1$D superlattice
chains~\cite{pai1,pai2,pai3,pai4,wang}, which are considered as the 
equivalent representation of periodic structures of different metallic 
layers~\cite{multi1,multi2,multi3}. The motivation of the present study is
two-fold. Firstly, we intend to reveal the interplay of the superlattice 
configuration in the ladder network and the external electric field on 
electronic conductance. If the mobility edge phenomenon in such a system
still persists at multiple electron energies like a $1$D superlattice chain 
in presence of external electric field, then a ladder network of this 
particular type, can be used in electronic circuits as a switching device 
which might throw new light in the present era of nanotechnology. Secondly,
the ladder networks are extremely suitable for explaining charge transport
through double stranded DNA~\cite{yi,diaz,apalkov}. A clear understanding 
of it is highly important due to its potential application in 
nano-electronics, especially, to design DNA-based computers. Not only that, 
it has also broad applications in biology since it is directly related to 
cell division, protein binding processes, etc. Experimental results suggest 
that DNA can have all possible conducting properties like, 
metallic~\cite{rakitin}, semiconducting~\cite{cai,porath} or even 
insulating~\cite{pablo,braun}. In spite of the numerous experimental and 
theoretical~\cite{th1,th2,th3} progress of charge transport through double 
helix DNA structures, the conducting nature of DNA molecules are not yet 
well explained and a deeper analysis is still needed, in the context of 
fundamental physics and technological interest. Nowadays scientists are 
trying to develop artificial DNA molecules~\cite{doi}, based on DNA 
synthesis equipment, those are highly stable and analogous to naturally 
occurred DNA molecules. Therefore, the study of electronic transport in a 
ladder network comprising of superlattices can give rise significant new 
results.

To make the present model much more realistic one should consider the 
effect of backbones and the environment. Though backbone sites are not 
directly involved on electronic conduction, but the effect of environment 
can be well analyzed since it essentially tunes the on-site potentials of 
these sites which results some interesting patterns. In our present work
we consider different arrangements of backbone sites (see Fig.1) and 
consider their effect on electronic conduction, to make it a self-contained
study.

The rest of the article is arranged as follows. In Section II we present
the model and the calculation method. The numerical results from our model 
calculations are illustrated in Section III, and finally, in Section IV 
we summarize the essential results.
 
\section{Model and theoretical framework}

Here we present the model quantum system considered for this work and the 
theoretical formulation to describe electronic transport properties in
\begin{figure}[ht]
{\centering \resizebox*{8.7cm}{2.2cm}{\includegraphics{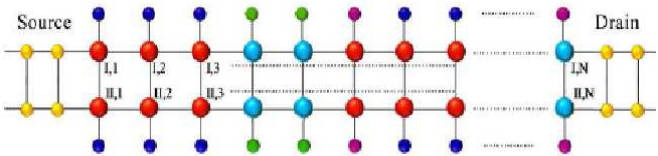}}\par}
\caption{(Color online). A tight-biding ladder network coupled to finite 
width source and drain electrodes. The filled circles with different colors 
correspond to different atomic sites.}
\label{ladder2}
\end{figure}
presence of an external bias. We use a simple TB framework to describe the 
system Hamiltonian, and, within non-interacting picture this approach is
highly suitable for analyzing electron transport through a bridge system.

\subsection{The TB model}

We consider a conducting junction in which a ladder network is clamped 
between source and drain electrodes. A sketch of such a 
source-conductor-drain junction is presented in Fig.~\ref{ladder2}, where
the filled circles with different colors correspond to different atomic
sites. The Hamiltonian of this full system can be written as sum of three
terms like,
\begin{equation}
\mbox{\boldmath$H$}=\mbox{\boldmath$H$}_{\mbox{\tiny lad}} + 
\mbox{\boldmath$H$}_{\mbox{\tiny eltd}} + 
\mbox{\boldmath$H$}_{\mbox{\tiny tun}}.
\label{hsum}
\end{equation}
The first term of Eq.~\ref{hsum} represents the Hamiltonian of the ladder
network, while the next two terms correspond to the Hamiltonians of the
side attached electrodes and electrode-to-ladder couplings, respectively.

Under nearest-neighbor hopping approximation the TB Hamiltonian of the
ladder network in site representation reads as,
\begin{eqnarray}
\mbox{\boldmath$H$}_{\mbox{\tiny lad}}&=&\sum_{i=I,II} \sum_j
\left[\epsilon_{i,j} \mbox{\boldmath$c$}_{i,j}^{\dagger} 
\mbox{\boldmath$c$}_{i,j} + t \left( \mbox{\boldmath$c$}_{i,j}^{\dagger} 
\mbox{\boldmath$c$}_{i,j+1} + \mbox{h.c.}\right) \right] \nonumber \\
& + & \sum_j v \left(\mbox{\boldmath$c$}_{I,j}^{\dagger}     
\mbox{\boldmath$c$}_{II,j} + \mbox{\boldmath$c$}_{II,j}^{\dagger}     
\mbox{\boldmath$c$}_{I,j} \right) \nonumber \\
& + & \sum_{i=I,II} \sum_j \left[\epsilon_{i,j}^b 
\mbox{\boldmath$c$}_{i,j}^{b\dagger} \mbox{\boldmath$c$}_{i,j}^b + 
t^b \left( \mbox{\boldmath$c$}_{i,j}^{b\dagger} \mbox{\boldmath$c$}_{i,j}^b 
+ \mbox{h.c.}\right) \right]
\label{hladder}
\end{eqnarray}
where the index $i$($=I,II$) corresponds to two strands of the ladder
and the index $j$($=1,2,3,\dots$) refers to the atomic sites in these 
strands. Thus, using the notation ($i,j$) we can locate atomic sites in
any strand. Each of these sites is again directly coupled to a backbone 
site (see Fig.~\ref{ladder2}) which essentially captures the effect of the 
environment. $\epsilon_{i,j}$ is the on-site energy of an electron at the
site $j$ of $i$-th strand, and $t$ and $v$ are the intra-strand and 
inter-strand hopping integrals, respectively.
$\mbox{\boldmath$c$}^{\dagger}_{i,j}$ ($\mbox{\boldmath$c$}_{i,j}$) is the 
creation (annihilation) operator of an electron at the site $(i,j)$. For 
the backbones we use an additional index $b$ in our TB model to describe 
the parameters like, on-site energies and hopping matrix elements, such 
that they are distinguished from the parameters used in the parent ladder. 
The backbone sites are modeled by the on-site energies $\epsilon_{i,j}^b$, 
while their couplings to the atomic sites of the strands are characterized 
by the nearest-neighbor hopping integrals $t^b$.

Depending on the arrangements of atomic sites in two different strands as 
well as in backbones, several cases are analyzed to describe the localization
phenomena in presence of an external voltage bias. We essentially focus on 
two different setups. In one case the upper and lower strands are arranged 
with superlattice configurations, setting the uniform site energies in the 
backbones, while in the other case the scenario gets reversed. Here, the 
strands are arranged with identical lattice sites and the backbone sites 
are configured with superlattice sites. In the present communication, we 
describe the superlattice models considering two or three types of atoms, 
in each unit cell, those are labeled as $\alpha$ and $\beta$ or $\alpha$, 
$\beta$ and $\gamma$, and they are arranged following a particular sequence. 
For example, it can be either $\alpha^p \beta^q$ or $\alpha^p \beta^q 
\gamma^r$, where $p$, $q$ and $r$ are three positive integers. Repeating
such unit cells we will get the entire lattice chain and construct the
desired full system for our study. 

In presence of a finite voltage bias $V$ between the source and drain
electrodes, an electric field is established which results the site
energies of the bridging conductor voltage dependent. Mathematically it 
reads as,
\begin{eqnarray}
\epsilon_{i,j}=\epsilon_{i,j}(0)+\epsilon_{i,j}(V)
\label{ss}
\end{eqnarray}
where, $\epsilon_{i,j}(0)$ represents the site energy in absence of external 
bias $V$. Depending on the nature of atomic sites $\epsilon_{i,j}(0)$'s are
called as $\epsilon_{\alpha}(0)$, $\epsilon_{\beta}(0)$, 
$\epsilon_{\gamma}(0)$ corresponding to $\alpha$, $\beta$, $\gamma$ type 
atomic sites, respectively. The voltage dependence in site energies 
essentially comes from two regions. One from the bare electric field 
\begin{figure}[ht]
{\centering \resizebox*{6cm}{3.5cm}{\includegraphics{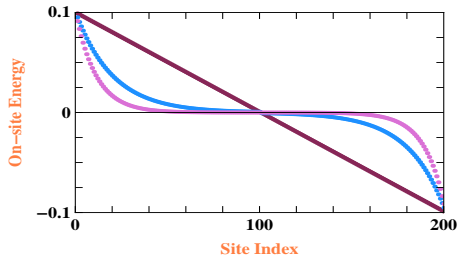}}\par}
\caption{(Color online). Voltage dependent site energies for three different 
potential profiles in a $200$-rung ladder network when the external bias $V$
is set at $0.2$.}
\label{profile}
\end{figure}
in the source-conductor-drain bias junctions and the other from the
screening in presence of long-range electron-electron (e-e) interactions, 
which is not directly taken into account in the Hamiltonian Eq.~\ref{hladder}.
If initially we don't consider the effect of such screening due to e-e
interactions, then the electric field becomes uniform along the ladder
and it gets the form,
\begin{equation}
\epsilon_{i,j}(V)=V/2 -\frac{j V}{N_r+1} 
\label{eqfld}
\end{equation}
where, $N_r$ is the total number of rungs in the ladder. Each of these 
rungs is attached to two backbone sites in opposite sides of the strands
($i=I,II$) whose site energies are also identically modified, following
Eq.~\ref{eqfld}, in presence of voltage bias like the parent lattice sites 
in individual rungs. Thus, site energies of four atomic sites (two parent
and two backbone sites) in each vertical line get equally modified in 
presence of external bias and their voltage dependence depend only on the
distance from the finite width source electrode which results Eq.~\ref{eqfld}
$i$ independent. Now, if we consider the screening effect, then the electric 
field will no longer be linear. Two such cases are shown in 
Fig.~\ref{profile} for two different screening strengths, as illustrative
examples. Below, we will analyze the localization phenomena in the ladder
network considering both linear and non-linear bias drops.

The TB Hamiltonians of the finite width electrodes 
($\mbox{\boldmath$H$}_{\mbox{\tiny eltd}}$) and electrode-ladder couplings
($\mbox{\boldmath$H$}_{\mbox{\tiny tun}}$) are also expressed quite similar
to Eq.~\ref{hladder}, where the electrodes are described with on-site
energies $\epsilon_0$ and nearest-neighbor interactions $t_0$. Two atomic
sites of the ladder are directly coupled to the atomic sites of the source
electrode where the coupling strength is described by $\tau_S$ and it is
$\tau_D$ for the drain-ladder coupling.

\subsection{Theoretical methods: Green's function approach}

To find electronic transmission coefficient for this bridge setup, we use
Green's function formalism~\cite{fisher,datta1,datta2} which is quite robust
compared to other existing theories available in literature. In this 
approach an infinite dimensional system (since electrodes are semi-infinite) 
gets effectively reduced to the dimension of a finite size conductor clamped 
between two electrodes. 

Using Fisher-Lee relation~\cite{fisher}, the two-terminal transmission 
probability can be written as,
\begin{eqnarray}
T= {\mbox{Tr}} \left[ \mbox{\boldmath$\Gamma$}_S 
\mbox{\boldmath$G$}^r_{\mbox{\tiny lad}} \mbox{\boldmath$\Gamma$}_D 
\mbox{\boldmath$G$}^a_{\mbox{\tiny lad}} \right]
\label{transm}
\end{eqnarray}
where, $\mbox{\boldmath$\Gamma$}_S$ and $\mbox{\boldmath$\Gamma$}_D$ are 
the coupling matrices and $\mbox{\boldmath$G$}^r_{\mbox{\tiny lad}}$ and 
$\mbox{\boldmath$G$}^a_{\mbox{\tiny lad}}$ are the retarded and advanced 
Green's functions of the ladder network, respectively. The single particle 
Green's function operator representing the complete system i.e., ladder
including source and drain electrodes, for an electron having energy $E$ 
is defined as,
\begin{equation}
\mbox{\boldmath$G$}=\left[ (E+i \eta) \mbox{\boldmath$I$}- \mbox{\boldmath$H$} 
\right]^{-1}
\end{equation}
where, $\eta \rightarrow 0^+$. $\mbox{\boldmath$H$}$ is the Hamiltonian of 
the full system and $\mbox{\boldmath$I$}$ denotes the identity matrix. 
Introducing the concept of self-energies due to source and drain, the 
problem of finding $\mbox{\boldmath$G$}$ in the full Hilbert space of 
$\mbox{\boldmath$H$}$ can eventually be mapped to a reduced Hilbert space
of the finite ladder, and the effective Green's function 
$\mbox{\boldmath$G$}_{\mbox{\tiny lad}}^r$ looks like,
\begin{eqnarray}
\mbox{\boldmath$G$}^r_{\mbox{\tiny lad}}=\left[(E+i\eta)\mbox{\boldmath$I$} 
-\mbox{\boldmath$H$}_{\mbox{\tiny lad}}-\mbox{\boldmath$\Sigma$}_S^r 
- \mbox{\boldmath$\Sigma$}_D^r \right]^{-1}.
\label{greendevice}
\end{eqnarray}
where, $\mbox{\boldmath$\Sigma$}_S$ and $\mbox{\boldmath$\Sigma$}_D$ are the 
contact self-energies used to describe the effect of couplings of the ladder
to the source and drain, respectively. A detailed derivations of these 
self-energies are available in Refs.~\cite{datta1,datta2,nikolic,we1}.

The coupling matrices $\mbox{\boldmath$\Gamma$}_S$ and
$\mbox{\boldmath$\Gamma$}_D$ corresponding to the couplings of the ladder
to the source and drain electrodes are directly associated with the 
self-energies and they are,
\begin{eqnarray}
\mbox{\boldmath$\Gamma$}_{S(D)}&=&i \left[ \mbox{\boldmath$\Sigma$}^r_{S(D)}
-\mbox{\boldmath$\Sigma$}^a_{S(D)}\right] \nonumber\\
&=& -2~ {\mbox{Im}}\left( \mbox{\boldmath$\Sigma$}^r_{S(D)}\right).
\end{eqnarray}
$\mbox{\boldmath$\Sigma$}^r_{S(D)}$, on the other hand, is the sum
$\mbox{\boldmath$\Sigma$}^r_{S(D)}=\mbox{\boldmath$\Lambda$}_{S(D)}
+i \mbox{\boldmath$\Delta$}_{S(D)}$, where the real part 
$\mbox{\boldmath$\Lambda$}_{S(D)}$ corresponds to the shift of energy 
levels of the ladder, and the imaginary part 
($\mbox{\boldmath$\Delta$}_{S(D)}$) is responsible for the broadening 
of these levels.

Finally, the average density of states (ADOS) of the ladder is determined 
from the relation,
\begin{equation}
\rho(E)=-\frac{1}{\pi Ns} {\mbox{Im}} \left[{\mbox{Tr}} \,
\mbox{\boldmath$G$}^r_{\mbox{\tiny lad}}\right]
\end{equation}
where, $N_s$ is the total number of atoms in the ladder network.

\section{Numerical Results and Discussion}

In this section, we describe numerical results obtained from the above
theoretical prescription given in Section II to investigate localization 
\begin{figure}[ht]
{\centering \resizebox*{8cm}{13cm}{\includegraphics{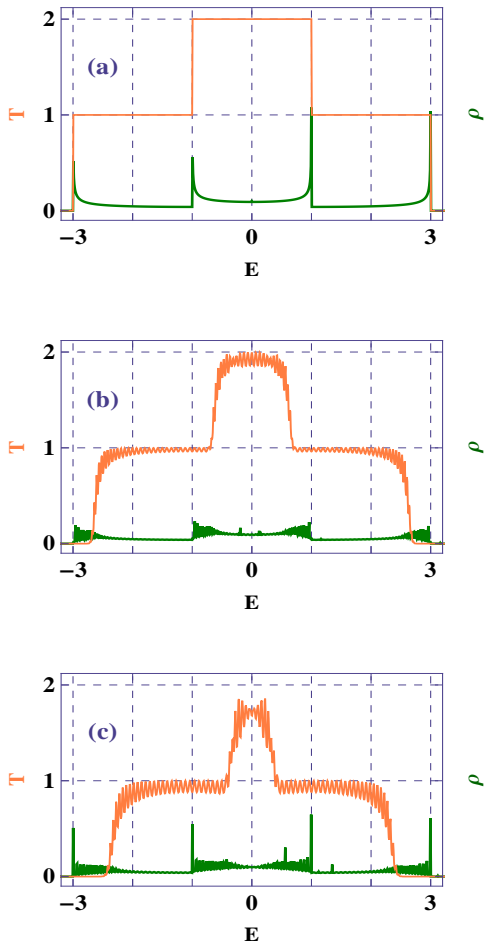}}\par}
\caption{(Color online). Two-terminal transmission probability $T$
(orange line) as a function of electron energy $E$ for a perfect ladder,
in absence of backbone sites, considering the linear potential drop as
shown by the red curve in Fig.~\ref{profile}, where (a), (b) and (c)
correspond to $V=0$, $0.2$ and $0.4$, respectively. The ADOS (green line)
is superimposed in each spectrum. Here we set $N_r=80$.}
\label{figvt}
\end{figure}
phenomena in the ladder network subjected to an external bias. Throughout 
the analysis we fix the electronic temperature at zero and choose $c=e=h=1$
for simplification. All the energies are scaled in unit of the hopping 
integral $t$.

As stated earlier, two different cases are analyzed depending on the 
superlattice configurations in the strands and backbones. In one setup,
the strands are arranged with superlattice sites considering identical
backbone sites, while in the other setup its opposite consequence is taken
into account. 
\begin{figure*}[t]
{\centering \resizebox*{17cm}{15cm}{\includegraphics{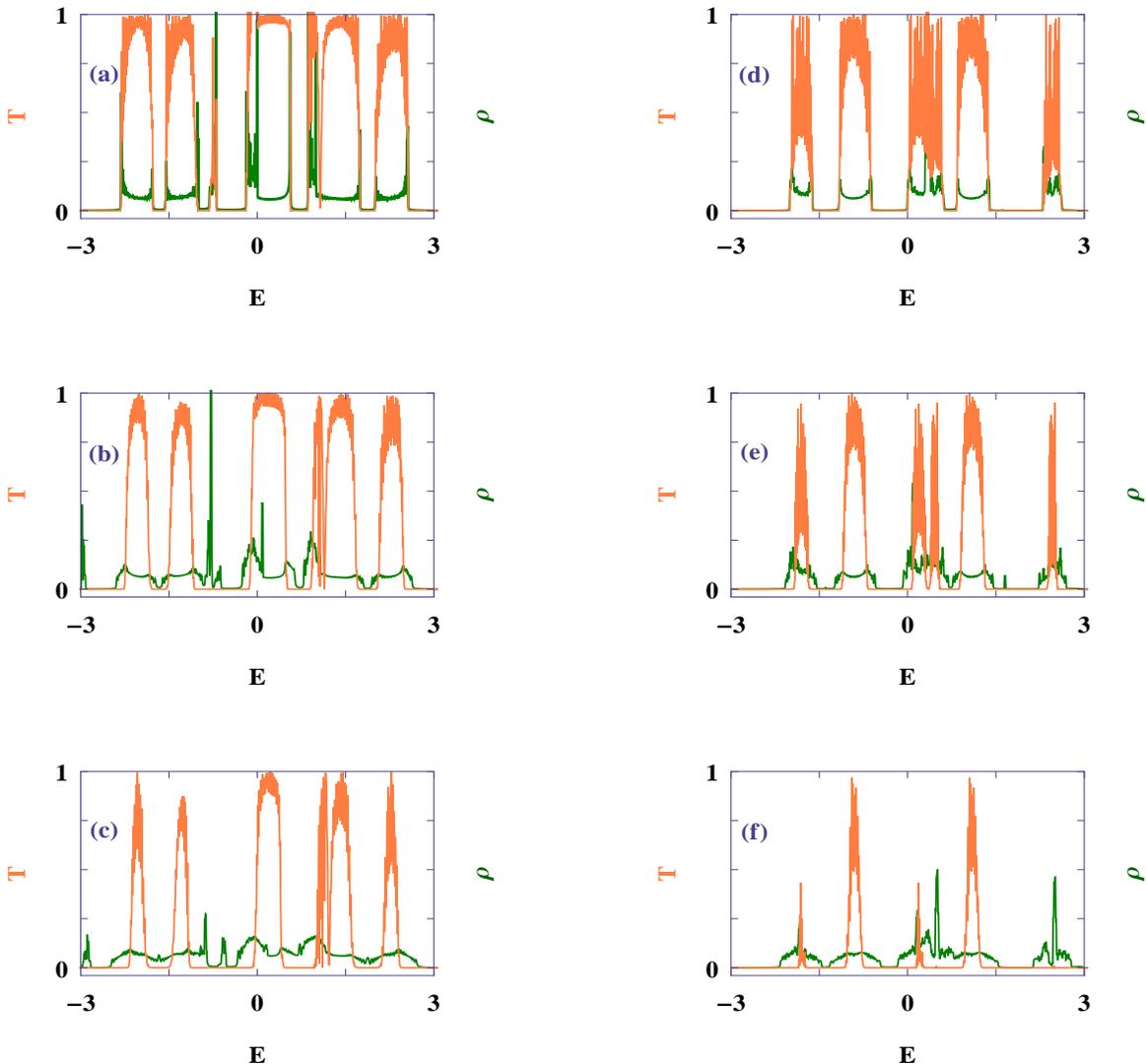}}\par}
\caption{(Color online). Two-terminal transmission probability $T$ together 
with average density of states $\rho$ as a function of energy $E$ for a
$150$-rung ladder network, considering of a linear bias drop along the 
ladder, where the upper, middle and lower rows correspond to $V=0$, $0.2$
and $0.4$, respectively. Two columns represent two different arrangements
of $\alpha$ and $\beta$ sites. In the left column, the results are computed
for the network where we choose $p=3$ and $q=2$ in the upper strand, and
$p=q=1$ in the lower strand. While, in the other column we set $p=3$ and
$q=2$ in both strands. All these results are done for $t^b=0$. The orange 
and green lines correspond to the identical meaning as stated in
Fig.~\ref{figvt}.}
\label{figord}
\end{figure*}
\begin{figure*}[t]
{\centering \resizebox*{17cm}{11cm}{\includegraphics{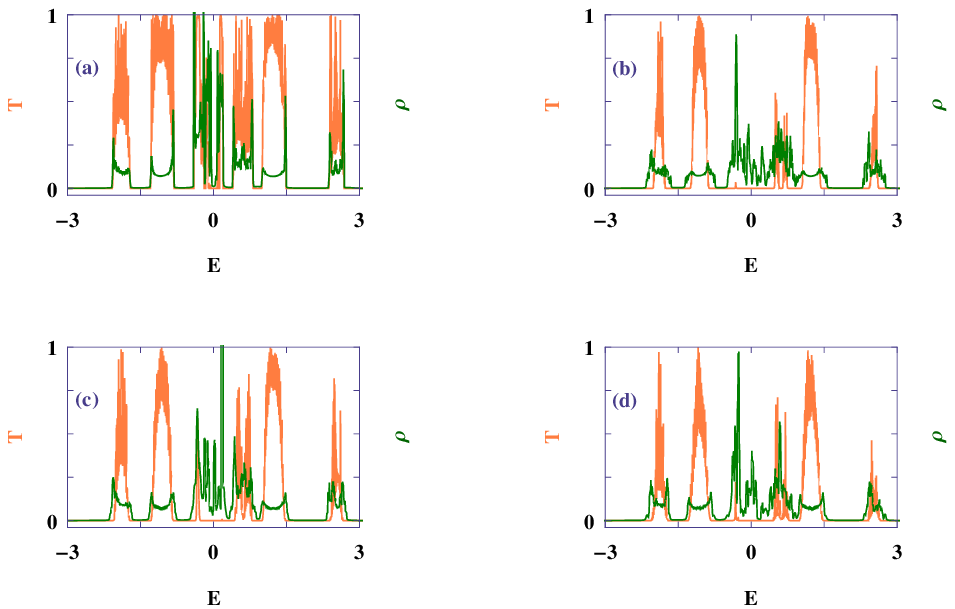}}\par}
\caption{(Color online). Two-terminal transmission probability and 
average density of states as a function of energy $E$ for a $150$-rung
ladder network considering $t^b=0.4$, where the orange and green curves
represent the similar meaning as stated earlier. The results shown in (a)
and (b) are computed for a linear potential profile, whereas (c) and (d)
are done for the non-linear potential profiles given by the blue and
pink lines in Fig.~\ref{profile}, respectively. In (a) we set $V=0$, 
while in the other three spectra (c-d) we choose $V=0.2$. All these results
are performed setting $p=3$ and $q=2$ in the upper and lower strands of the
ladder network.}
\label{figtb}
\end{figure*}

Before addressing these central cases, let us first discuss the effect 
of voltage bias on electron transport in a simple system which is a perfect
ladder without any backbone sites. The results of such a simple model is
presented in Fig.~\ref{figvt}, where we choose $N_r=80$. For this ladder, 
the voltage independent sites energies ($\epsilon_{i,j}(0)$'s) are identical, 
and therefore, they can be fixed at zero, without any loss of generality. 
The other parameters used here are as follows. The nearest-neighbor hopping 
integrals $t$ and $v$ are fixed at $1$, and the ladder-electrode coupling 
constants are taken to be $\tau_S=\tau_D=1$. In side-attached electrodes, 
the on-site potentials $\epsilon_0$ and hopping integrals $t_0$ are set at 
$0$ and $1$, respectively. Depending on the potential drop between the 
source and drain electrodes, three different cases are taken into account
those are presented in Figs.~\ref{figvt}(a), (b) and (c) where we select 
$V=0$, $0.2$ and $0.4$, respectively. In each of these three spectra 
two-terminal transmission probability together with average density of
states are shown. All these results are computed considering a linear 
potential profile along the ladder. In the absence of any electric field
associated with the voltage bias, electrons can conduct through the ladder 
for all possible allowed energies corresponding to the energy eigenvalues 
of the ladder which can be clearly noticed from the upper panel of 
Fig.~\ref{figvt}(a). Under this condition i.e., $V=0$, on-site potentials 
are no longer affected and it results extended energy levels throughout 
the energy band window of the ordered ladder. The situation becomes 
somewhat interesting when a finite potential drop appears along the ladder
(Figs.~\ref{figvt}(b) and (c)). It is observed that several energy levels 
along the energy band edges do not contribute anything to the electron
conduction i.e., transmission coefficient drops to zero for these energies.
Obviously, if the Fermi energy of the ladder lies within these regions,
no electron transmission takes place and the insulating phase is obtained.
The finite transmission will take place only when the energy moves towards
the band centre. Thus, a sharp transition between a fully insulating zone
to a conducting zone is obtained which gives rise to the phenomenon of 
mobility edge in presence of external electric field. In presence of a 
non-zero bias, site energies are voltage dependent (see Eq.~\ref{ss}), 
those are also not identical to each other, which are responsible for 
generating localized energy levels in the energy spectrum. Certainly, more 
such localized energy levels will be available for higher potential drop. 
This is exactly shown in Fig.~\ref{figvt}(c), considering a finite bias 
$V=0.4$. It should be noted that, like a conventional disordered material, 
the localization of energy levels always starts from the band edges, keeping 
the extended states towards the band centre. This phenomenon has been 
revisited in our recent work in the context of studying integer quantum Hall 
effect~\cite{hall} within a tight-binding framework. It reveals that, for 
large enough electric field almost all the energy levels get localized, and 
therefore, no such crossover between a conducting zone and an insulating 
phase is observed. 

Based on the above analysis of electronic localization in a perfect ladder
in presence of external electric field now we concentrate on the central
results i.e., the interplay of superlattice configuration in parent strands
as well as in backbones of the ladder network and the external electric
field associated with the voltage bias. In what follows, we present our
results in two sub-sections for two different setups.

\subsection{The TB ladder network with superlattice chains and identical
backbone sites}

Here we present our numerical results for the ladder network considering 
superlattice sites in the strands, setting identical lattice sites in
backbones. Two different types of atomic sites, labeled as $\alpha$ and 
$\beta$, are taken into account and depending on their arrangements in 
the strands several cases are analyzed. These arrangements are specified
by the rule $\alpha^p\beta^q$, as stated earlier, where $p$ and $q$ are 
variables. The parameters used here for the calculations, unless stated 
otherwise, are $\epsilon_{\alpha}(0)=-\epsilon_{\beta}(0)=1$, 
$\epsilon_{\alpha}^b(0)=\epsilon_{\beta}^b(0)=0$, $t=v=1$, $\epsilon_0=0$, 
$t_0=1$ and $\tau_S=\tau_D=1$. The hopping integral $t^b$ is mentioned
in the figure captions.

As illustrative examples, in Fig.~\ref{figord} we present two-terminal
transmission probability together with average density of states of
\begin{figure*}[t]
{\centering\resizebox*{17cm}{15cm}{\includegraphics{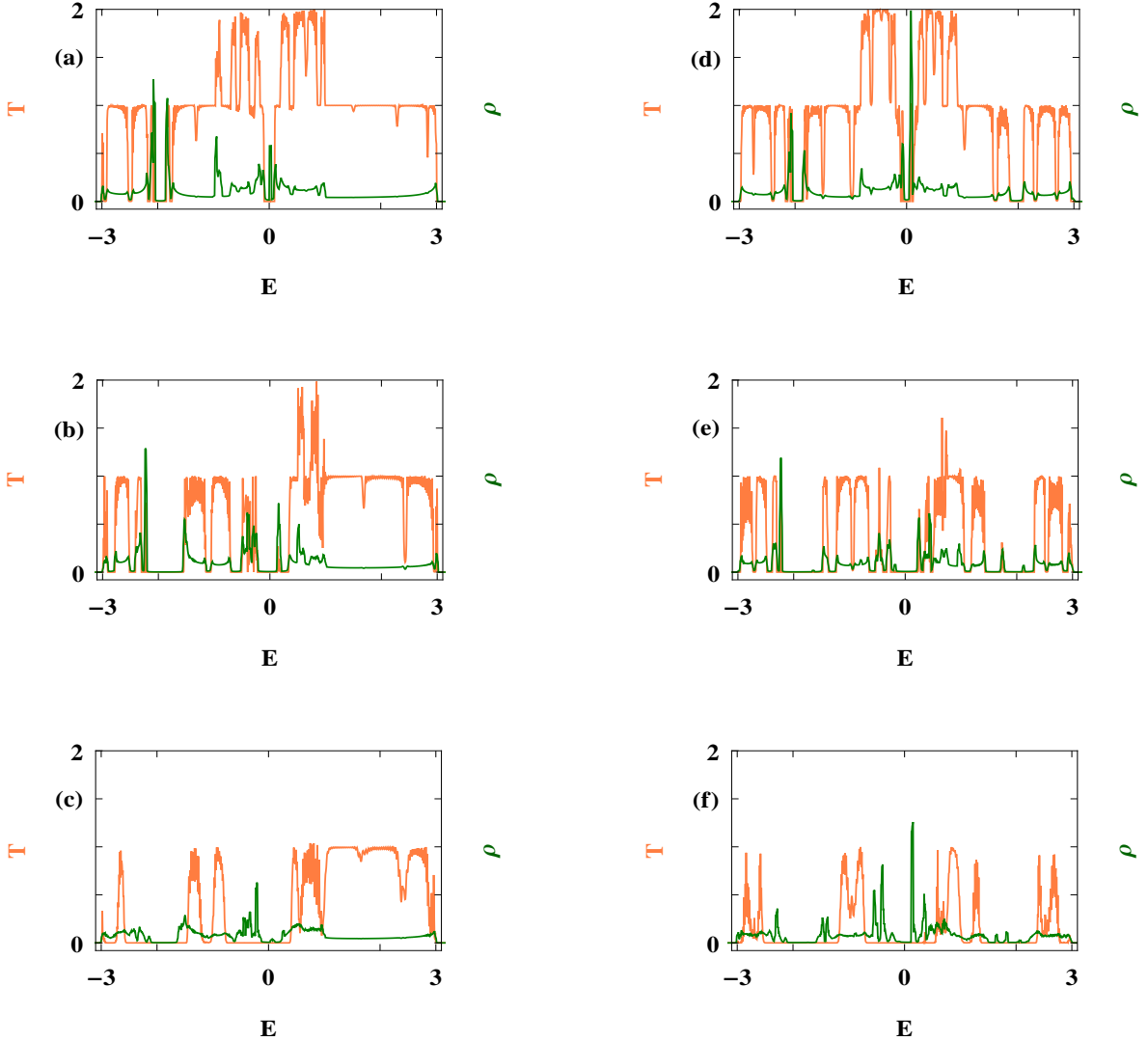}}\par}
\caption{(Color online). Two-terminal transmission probability $T$ and 
average density of states $\rho$ as a function of energy $E$ for a ladder 
network in which strands are arranged with identical lattice sites and the 
backbones are configured with superlattice sites. In the left column,
backbones are arranged with $\alpha$ and $\beta$ sites where we set $p=5$ 
and $q=2$, while in the right column they are arranged with $\alpha$,
$\beta$ and $\gamma$ sites considering $p=5$, $q=3$ and $r=2$. The results
given in the 1st row are done with $t^b=0.4$, while all the other results
are computed when $t^b$ is fixed at $0.8$. The voltage $V$ is taken to be
zero in the 1st and 2nd rows, whereas it is $0.2$ in the 3rd row. All these
results are performed considering a linear electrostatic potential profile.
Moreover, we choose total number of rungs $N_r=140$ for the left column,
while it is $150$ in the right column. Two different colors correspond to 
the identical meaning as mentioned in the above figures.}
\label{backbone}
\end{figure*}
a $150$-rung ladder network, considering $t^b=0$ i.e., effect of backbone
sites is ignored, for different configurations of $\alpha$ and $\beta$ sites 
in two strands. For example, in the left column the results are shown when 
we choose $p=3$ and $q=2$ in the upper strand and these parameters are set 
equal to $1$ for the other strand. On the other hand, in the right column 
we select $p=3$ and $q=2$ for both these two strands. All these results are 
computed taking a linear potential profile along the ladder. Quite 
interestingly we see that, in presence of superlattice sites multiple
energy bands, separated by finite gaps, are appeared in the energy spectra. 
Depending on the arrangements of superlattice sites, i.e., the choices of
unit cells in two different strands, total number of such energy bands 
varies. Under this situation when we consider the effect of external 
electric field associated with the voltage bias, we get multiple localized 
regions, generated from the band edges, those are separated from the extended 
energy zones (see the spectra in Fig.~\ref{figord}). Therefore, setting
the Fermi energy in suitable energy regions the ladder network can be
used to transmit electrons or not from the source to drain electrode, and
thus, support in fabricating mesoscopic switching devices at multiple
energies. This essentially leads to the phenomenon of getting multiple
mobility edges.
 
The appearance of energy bands, separated by finite energy gaps, in presence 
of superlattice sites in the strands is strongly affected by the backbone 
sites, and the localization properties for a particular configuration, on
the other hand, are highly influenced by the nature of electrostatic 
potential profiles. These issues are addressed in Fig.~\ref{figtb}, where 
we choose $N_r=180$ and set $p=3$ and $q=2$ in both strands of the ladder
network. In the upper panel the results are calculated considering a 
linear bias drop along the sample, while in the lower panel they are 
computed for non-linear potential profiles. Comparing the spectra shown 
in Figs.~\ref{figord}(d) and \ref{figtb}(a) (since in these two cases
all other parameters, except $t^b$, are identical) we emphasize that the
ADOS spectra can be controlled in a significant way by means of the 
backbone environment. This change in the density profile becomes much 
pronounced when the difference between the parameters $p$ and $q$ is 
quite large, which we justify through our extensive numerical analysis. 
The effects of backbone sites is elaborately discussed in the following
sub-section.

Keeping in mind the possibilities of having much more electron screening
now we present the results (Figs.~\ref{figtb}(c) and (d)) where 
non-linear potential profiles are considered. With increasing the flatness
of the electrostatic potential profile, the localized regions get 
decreased. Eventually, for the extreme condition i.e., when the bias
drop takes place only at the ladder-to-electrode interfaces no such 
localized regions will be obtained, and certainly, metal-to-insulator
transition cannot be obtained.

So far the results discussed are for the systems where strands are arranged
with superlattice configuration. Now the question is, can one get multiple
mobility edges if the backbone sites are modulated with superlattice sites 
instead of the strands? If yes, then few more advantages are there than the
cases studied above. First, one can easily tune backbone site energies by
placing the sample in the vicinity of external gate electrodes which can
control the band spectra, and thus, results a selective switching action 
for a fixed Fermi energy in presence of external electric field. Second, 
the environmental effect on electron transport as present in natural DNA
molecules can also be explained through our model quantum systems. Below,
these issues are discussed with some typical examples.

\subsection{The TB ladder network with identical lattices in strands and 
superlattice sites in backbones}

Finally, we consider the system where strands are arranged with identical
atoms and backbones are arranged with superlattice sites. Some typical
cases are analyzed depending on the choices of superlattice sites and the
results are presented in Fig.~\ref{backbone}. For example, in the left column 
of Fig.~\ref{backbone} the backbone sites are arranged with two different
atomic sites $\alpha$ and $\beta$ obeying the sequence $\alpha^p\beta^q$,
while in the right column three atomic sites $\alpha$, $\beta$ and $\gamma$
are used to construct the backbones following the relation 
$\alpha^p\beta^q\gamma^r$. The parameters considered for these calculations
are: $\epsilon_{\alpha}(0)=\epsilon_{\beta}(0)=\epsilon_{\gamma}(0)=0$,
$\epsilon_{\alpha}^b(0)=-2$, $\epsilon_{\beta}^b(0)=0$, 
$\epsilon_{\gamma}^b(0)=2$, $t=v=1$, $\epsilon_0=0$, $t_0=1$ and
$\tau_S=\tau_D=1$. The hopping strength $t^b$ is given in the figure
caption. From the spectra (Fig.~\ref{backbone}), it is interesting to 
note that multiple energy bands separated by finite gaps are also
obtained here like the cases for the ladders with superlattice strands.
At the same time, the gap between the energy bands can be tuned by 
regulating the coupling strength $t^b$ which may provide additional
control parameter for getting selective switching effect at the meso-scale
level for multiple energies.

\section{Concluding Remarks}

In the present communication, we address electric field effect associated
with the applied voltage bias on localization phenomena in a tight-binding 
ladder network with superlattice configuration. Using Green's function 
approach we compute two-terminal transmission probability along with average 
density of states for different arrangements of superlattice sites in the 
strands as well as in the backbones. Our numerical results exhibit several 
interesting features which essentially lead to a possibility of using the 
ladder network as a switching device at multiple energies. This switching
action is significantly controlled by many factors like, the nature of
electrostatic potential profile, presence of backbone sites, coupling
between the backbone sites and the sites in the parent strands, etc.
With increasing the flatness of the potential profile, related to screening
effects, localized energy regions across the band edges gradually decreases,
and, for the typical case when potential drop takes place only at the 
electrode-ladder interfaces, localized regions disappear, and thus, 
metal-to-insulator transition cannot be observed upon changing the
system Fermi energy. Similarly, the backbone sites are also responsible
for controlling the switching action which practically suggest us an
experiment towards this direction.

Finally, it is important to state that, although the electronic temperature
is fixed at zero throughout the analysis, all these results are still valid
in the low-temperature limit. This is due to the fact that the thermal
broadening of energy levels is much weaker than the broadening caused by 
ladder-to-electrode coupling~\cite{datta1,datta2,we2,we3}.

\section*{Acknowledgment}

Second author is thankful to Prof. A. Nitzan for useful conversations.

\end{document}